\documentclass{emulateapj}
\usepackage{longtable}
\usepackage{epsfig}
\usepackage{lscape}

\newcommand {\apgt} {\ {\raise-.5ex\hbox{$\buildrel>\over\sim$}}\ }
\newcommand {\aplt} {\ {\raise-.5ex\hbox{$\buildrel<\over\sim$}}\ }

\newcommand{\etal}{et al.~}%

\def\sun{\hbox{$\odot$}}
\def\cm3{cm$^{-3}$}

\def\rsun{R$_{\odot}$}

\def\msun{M$_{\odot}$}

\def\beq{\begin{equation}}
\def\eeq{\end{equation}}

\def\sles{\lower2pt\hbox{$\buildrel {\scriptstyle <}
   \over {\scriptstyle\sim}$}}
\def\sgreat{\lower2pt\hbox{$\buildrel {\scriptstyle >}
   \over {\scriptstyle\sim}$}}

\slugcomment{Accepted to the Astrophysical Journal}
\shorttitle{Ultra-Bright Type II-L SN 2008es}
\shortauthors{Gezari et al.}

\begin{document}

\title{DISCOVERY OF THE ULTRA-BRIGHT Type II-L SUPERNOVA 2008\MakeLowercase{es}}

\author{S.~Gezari,\altaffilmark{1} J.~P.~Halpern,\altaffilmark{2} D.~Grupe,\altaffilmark{3} F.~Yuan,\altaffilmark{4} R.~Quimby,\altaffilmark{5} T.~McKay,\altaffilmark{4} D.~Chamarro,\altaffilmark{4} M.~D.~Sisson,\altaffilmark{4} C.~Akerlof,\altaffilmark{4} J.~C.~Wheeler,\altaffilmark{6} P.~J.~Brown, \altaffilmark{3} S.~B.~Cenko,\altaffilmark{7} A.~Rau,\altaffilmark{5} J.~O.~Djordjevic,\altaffilmark{8} and D.~M.~Terndrup\altaffilmark{8} 
}
\altaffiltext{1}{Department of Physics and Astronomy,
        Johns Hopkins University,
        3400 North Charles Street,
        Baltimore, MD 21218 \email{suvi@pha.jhu.edu}}

\altaffiltext{2}{Department of Astronomy,
     Columbia University,
         New York, NY  10027}

\altaffiltext{3}{Pennsylvania State University, Department of Astronomy \& Astrophysics, University Park, PA  16802}

\altaffiltext{4}{University of Michigan, Randall Laboratory of Physics,
        450 Church St., Ann Arbor, MI, 48109-1040}

\altaffiltext{5}{Division of Physics, Mathematics and Astronomy, 
California Institute of Technology,
        Pasadena, CA  91125}

\altaffiltext{6}{Department of Astronomy, University of Texas,
          Austin, TX  78712}

\altaffiltext{7}{Department of Astronomy, 601 Campbell Hall,
          University of California, Berkeley, CA  94720-3411}

\altaffiltext{8}{Department of Astronomy, Ohio State University, 
           Columbus, OH  43210}

\begin{abstract}

We report the discovery by the Robotic Optical Transient Search Experiment (ROTSE-IIIb) telescope of SN 2008es, an overluminous supernova (SN) at $z=0.205$ with a peak visual magnitude of $-$22.2.  We present multiwavelength follow-up observations with the \textsl{Swift} satellite and several ground-based optical telescopes. The ROTSE-IIIb observations constrain the time of explosion to be 23$\pm$1 rest-frame days before maximum.  The linear decay of the optical light curve, and the combination of a symmetric, broad H$\alpha$ emission line profile with broad P Cygni H$\beta$ and Na I~$\lambda5892$ profiles, are properties reminiscent of the bright Type II-L SNe 1979C and 1980K, although SN 2008es is greater than 10 times more luminous.  The host galaxy is undetected in pre-supernova Sloan Digital Sky Survey images, and similar to Type II-L SN 2005ap (the most luminous SN ever observed), the host is most likely a dwarf galaxy with $M_{r} > -17$.  \textsl{Swift} Ultraviolet/Optical Telescope observations in combination with Palomar 60 inch photometry measure the spectral energy distribution of the SN from 200 to 800 nm to be a blackbody that cools from 14,000 K at the time of the optical peak to 6400 K 65 days later.  The inferred blackbody radius is in good agreement with the radius expected for the expansion speed measured from the broad lines (10,000 km s$^{-1}$). The bolometric luminosity at the optical peak is $2.8 \times 10^{44}$ erg s$^{-1}$, with a total energy radiated over the next 65 days of $5.6 \times 10^{50}$ erg.  The exceptional luminosity of SN 2008es requires an efficient conversion of kinetic energy produced from the core-collapse explosion into radiation.   We favor a model in which the large peak luminosity is a consequence of the core-collapse of a progenitor star with a low-mass extended hydrogen envelope and a stellar wind with a density close to the upper limit on the mass-loss rate measured from the lack of an X-ray detection by the \textsl{Swift} X-Ray Telescope.  

\end{abstract}


\section{Introduction \label{intro}}
Hydrogen-rich supernovae (SNe II) produce their radiative energy in several phases; with a diversity of luminosities, light curves, and spectroscopic features that divide them into subclasses.  An initial burst of UV/X-ray radiation is observed at the time of shock breakout (Schawinski \etal 2008; Gezari \etal 2008a), followed by declining UV/optical emission from the adiabatic expansion and cooling of the SN ejecta (e.g., SN II-pec 1987A, Hamuy \etal 1988; SN IIb 1993J, Schmidt \etal 1993; Richmond \etal 1994, SN II-P 2006bp, Immler \etal 2007).  Type II-plateau (II-P) SNe are characterized by their subsequent plateau in optical brightness, which is understood as the result of a progenitor with a substantial hydrogen envelope, for which a cooling wave of hydrogen recombination recedes through the inner layers of the ejecta (Falk \& Arnett 1977; Litvinova \& Nadyozhin 1983).  Type II-linear (II-L) SNe do not have a plateau phase, but rather a steep drop in luminosity that is thought to be the result of the ejection of large amounts of radioactive material (Young \& Branch 1989) or a small hydrogen envelope mass (Barbon \etal 1979; Blinnikov \& Bartunov 1993).  The subclass of bright SNe II-L ($M_{B} < -18$; Patat \etal 1994) has been proposed to be the result of an extended low-mass envelope (Swartz \etal 1991) and reprocessing of UV photons in the superwind of the progenitor star (Blinnikov \& Bartunov 1993) or a gamma-ray burst (GRB) explosion buried in a hydrogen-rich envelope (Young \etal 2005).  Type II-narrow (IIn) SNe show signs of interaction of the ejecta with  circumstellar material (CSM) in the form of narrow emission lines, X-ray emission, and an excess in optical luminosity (e.g., SN 1988Z: Turatto \etal 1993).  After the photospheric phase in SNe II, a final exponential decline is powered by heating due to the radioactive decay of $^{56}$Co. 

In this paper, we present the discovery of an overluminous SN II that has the properties of an SN II-L, but with an exceptional luminosity that we argue is best explained by the core-collapse of a progenitor star with a low-mass, extended hydrogen envelope and a dense stellar wind.  We disfavor more exotic scenarios such as an interaction with a massive shell of CSM expelled in an episodic eruption, a pair-instability explosion, or a buried GRB. In \S \ref{sect_obs}, we present the Robotic Optical Transient Experiment (ROTSE-IIIb) discovery data, and follow-up observations with the \textsl{Swift} satellite and several ground-based telescopes; in \S \ref{sec_disc} we compare the properties of SN 2008es to other known Type II SNe, and use our observations to constrain the mechanism powering the extraordinary radiative output; and in \S \ref{sec_conc} we make our conclusions.

\section{Observations} \label{sect_obs}

\subsection{Photometry} \label{sect_optphot}

\begin{figure*}
\plotone{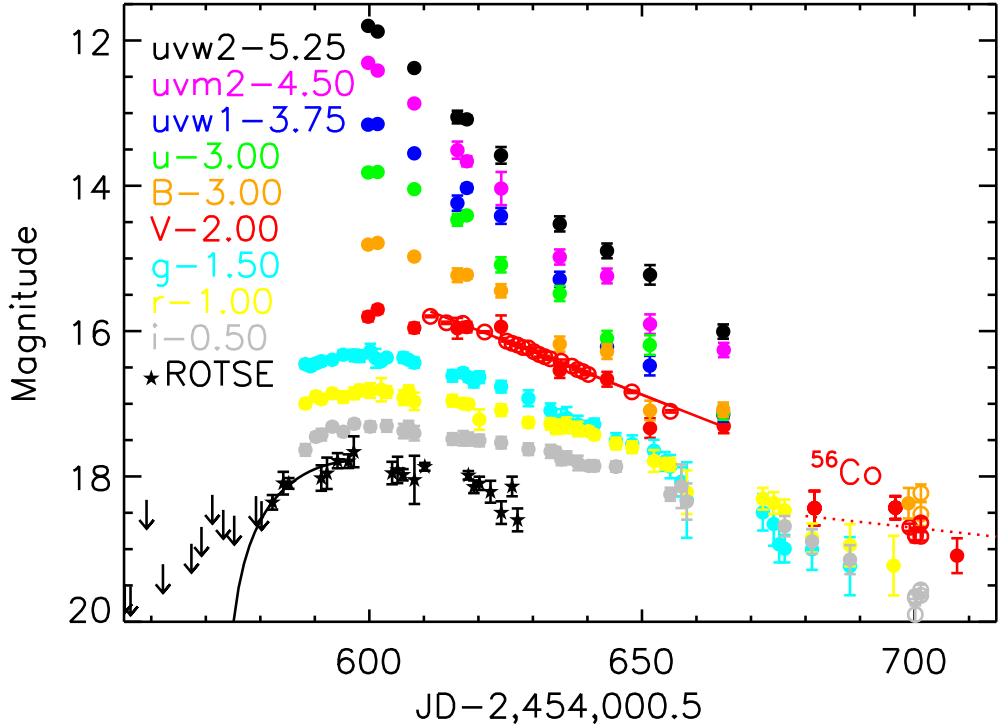}
\caption{
Light curve of SN 2008es in the unfiltered ROTSE-IIIb optical band (calibrated to the SDSS $r$ band), the Palomar 60 inch $g$, $r$, and $i'$ bands, the $V$ band from the MDM 1.3m (plotted with open circles), the $B$, $V$, and $i$-band data from P200 (plotted with open circles), and the $uvw2$, $uvm2$, $uvw1$, $u$, $B$, and $V$ bands measured by the \textsl{Swift} UVOT telescope.  The black line shows a quadratic fit to the rise in flux measured by ROTSE-IIIb used to determine the time of explosion at MJD $54,574\pm1$.  The dotted red line shows the slope of radioactive $^{56}$Co decay of 0.98 mag (100 days)$^{-1}$, stretched in time to match the time dilation of SN 2008es.
\label{fig_lc}
}
\end{figure*}

SN 2008es was discovered as an optical transient at $\alpha = $11h 56m 49$\fs$13, $\delta = +54^{\circ}$ 27$\arcmin$ 25$\farcs$7 (J2000.0) in unfiltered images (calibrated to the SDSS $r$ band) taken by the 0.45\,m ROTSE-IIIb telescope \citep{2003PASP..115..132A} at McDonald Obseratory on 2008 Apr 26 (ROTSE J115649.1+542726; Yuan \etal 2008).  UT dates are used throughout this paper.  ROTSE-IIIb continued to monitor the transient for $\sim 50$ days.  We began a monitoring program with the Palomar 60 inch (P60) telescope \citep{2006PASP..118.1396C} in the $g$, $r$, and $i'$ bands on 2008 May 2, with continued observations for $\sim 100$ days.  Observations in the $V$ band were obtained by the 1.3m MDM telescope and the 4 K detector with 0.315 arcsec pixel$^{-1}$ spatial resolution, starting on 2008 May 24 and continuing for 45 days.  Secondary standard stars were established through observations of stars from \citet{1992AJ....104..340L}, and included the determination of nightly extinction corrections.  Observations on 2008 August $22-23$ in the $B$, $V$, and $i$ bands were obtained with the Palomar 200 inch telescope (P200) and the Large Format Camera.  

The optical light curve measured from all four optical telescopes is shown in Figure \ref{fig_lc} and Table \ref{tab_1}.  Magnitudes reported in this paper have been corrected for a Galactic extinction of E($B-V$)=0.012\,mag \citep{1998ApJ...500..525S}.   The ROTSE-IIIb and P60 data catch the rise of the source to its peak.  We fit the rise of the unfiltered flux measured by ROTSE IIIb with a quadratic function (plotted with a black line in Figure \ref{fig_lc}), and estimate the time of explosion from the intercept with the time axis to be MJD 54,574 $\pm$ 1.  The error in the time of explosion reflects both the statistical error from the ROTSE-IIIb flux uncertainties and the systematic error from the shape of the function used for the fit.  We measure the peak of the light curve from quadratic fits to the individual filter light curves from MJD 54,580 to 54,620, which yields a peak of MJD 54,602 $\pm 1$, and corresponds to 23 $\pm$ 1\,days after the time of explosion in the rest frame of the SN.  

\subsection{Spectroscopy} \label{sect_optspec}
\begin{figure*}
\plotone{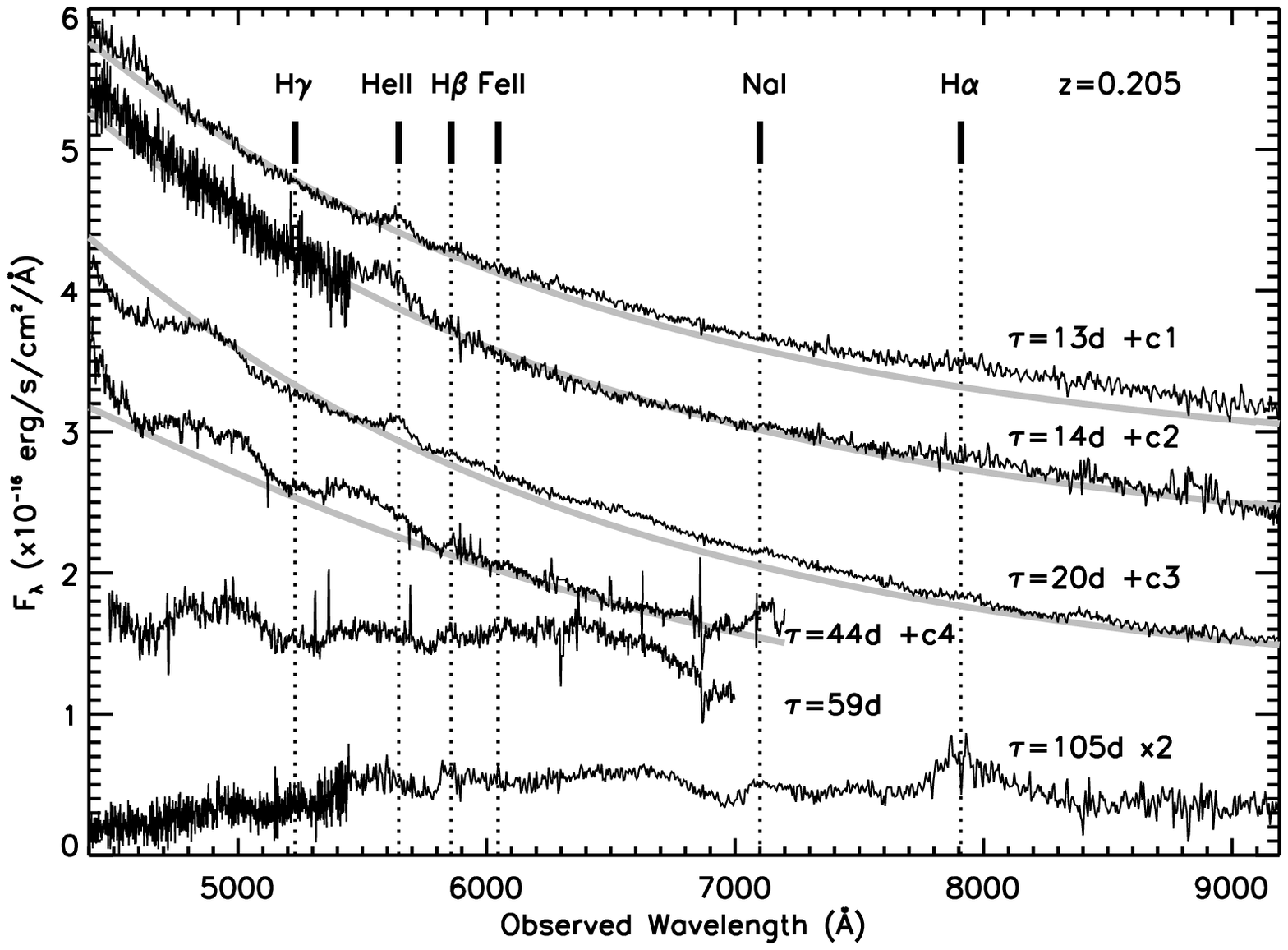}
\caption{
Optical spectra of SN 2008es taken on 2008 May 1 (day 13), May 8 (day 20), June 1 (day 44), and June 16 (day 59) with the HET 9.2m, and on 2008 May 2 (day 14), and August 1 (day 105), on the Palomar 200 inch, where $\tau$ is the observed days since explosion.  Spectra are offset in the vertical direction by $c1$ = 2.6, $c2$ = 2.0, $c3$ = 1.0, and $c4$=0.5, and the 2008 August 1 spectrum (day 105) is multiplied by a factor of 2 for comparison.  The redshift of $z=0.205$ is measured from the broad He II $\lambda$4686 line in the May 1 and 8 HET spectra.   The solid grey lines show a fit to the continuum with a blackbody at $z=0.205$ on May 1,2, and 8 with $T_{bb} = 14,000$ K and on June 1 with $T_{bb} = 11,000$ K.  
\label{fig_spec}
}
\end{figure*}

Within days of the ROTSE discovery, we followed up the source with optical spectroscopy on 2008 May 1 and 8 with the 9.2m Hobberly Eberly Telescope (HET) and the Marcario Low-Resolution Spectrograph and on 2008 May 2 with P200 and the Double Beam Spectrograph \citep{2008ATel.1515....1Y}.  The transient was tentatively classified as a quasar because the spectra showed a featureless blue continuum and one broad emission feature that was associated with Mg II $\lambda$2798 at $z=1.02$.  We obtained further follow-up spectra on 2008 June 1 and 16 with the HET and on 2008 August 1 with P200.  The identification of the redshift and type of the SN was ambiguous (Miller \etal 2008a; Gezari \etal 2008b; Miller \etal 2008b) until later spectra obtained on 2008 August 1 (by our group with P200) and 2008 August 3 (by Chornock \etal 2008 with the 10m Keck Low-Resolution Imager and Spectrograph), revealed broad Balmer emission lines, reported by Chornock \etal (2008) to be associated with a Type II SN at $z=0.21$.  Figure \ref{fig_spec} shows the evolution of the spectra over time, and Table \ref{tab_2} gives the log of spectroscopic observations.   In order to minimize uncertainties in the absolute flux calibration, the spectra are multiplied by a scaling factor so that their flux integrated over the $r$-band filter matches the P60 $r$-band photometry closest in time to the observation.  We see no evidence of narrow absorption or emission lines superimposed on the SN spectrum indicating a contribution from the host galaxy.

We measure the redshift of the SN to be $z=0.205 \pm 0.001$ from a Gaussian fit to the broad He II $\lambda 4686$ line in the 2008 May 1 and 8 spectra, with a full-width at half-maximum (FWHM) of 5500 $\pm~500$ km s$^{-1}$.   Detection of the high ionization line He II $\lambda 4686$ at $t = $14 to 20 days after explosion in the SN rest frame requires high temperatures and ionization.  This is similar to the case of Type II-P SN 2006bp for which He II $\lambda 4686$ was detected in the first few days after explosion, and was associated with outer layers of the SN ejecta that were flash ionized at shock breakout \citep{2007ApJ...666.1093Q,2008ApJ...675..644D}.  

\begin{figure*}
\plottwo{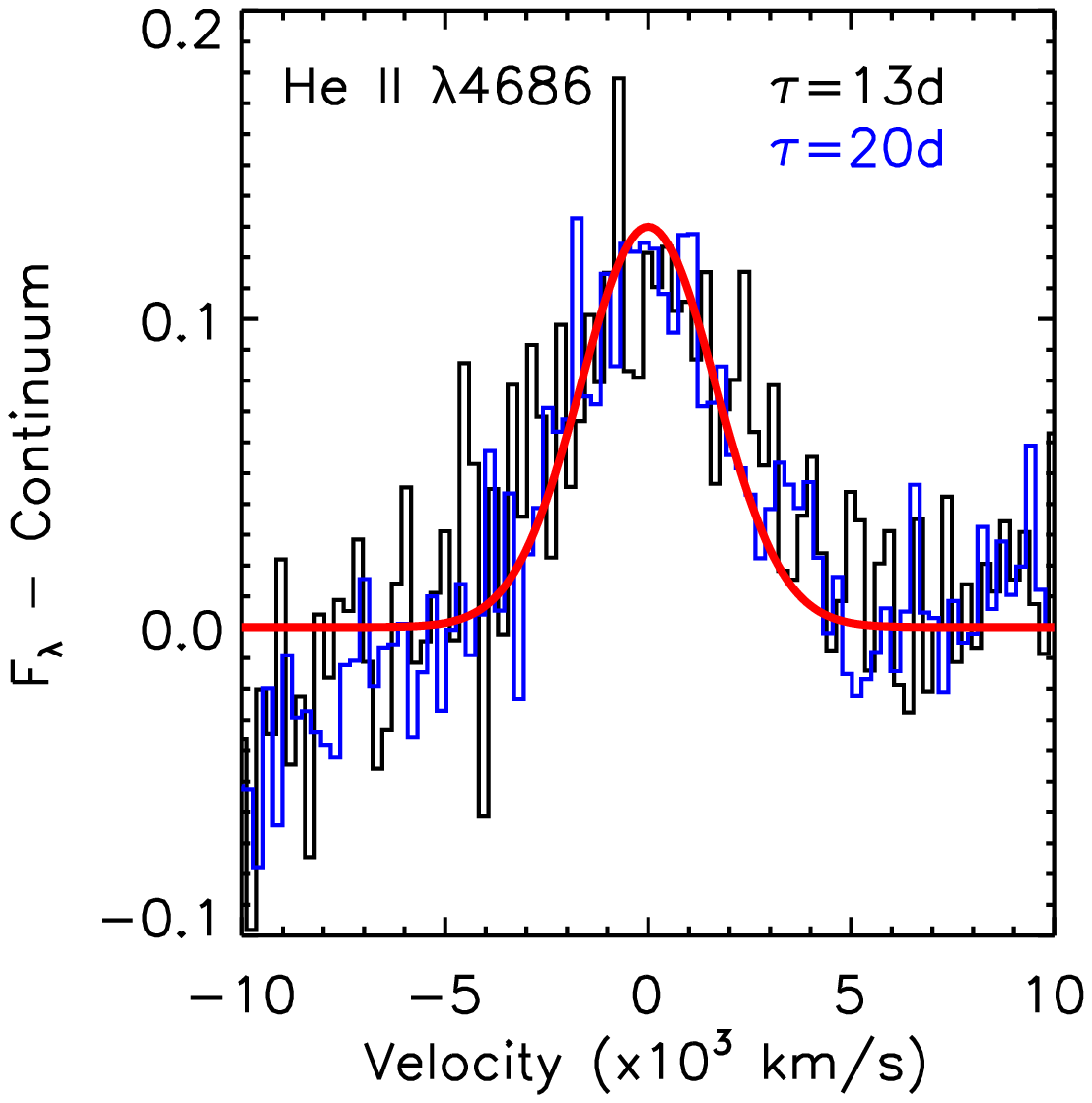}{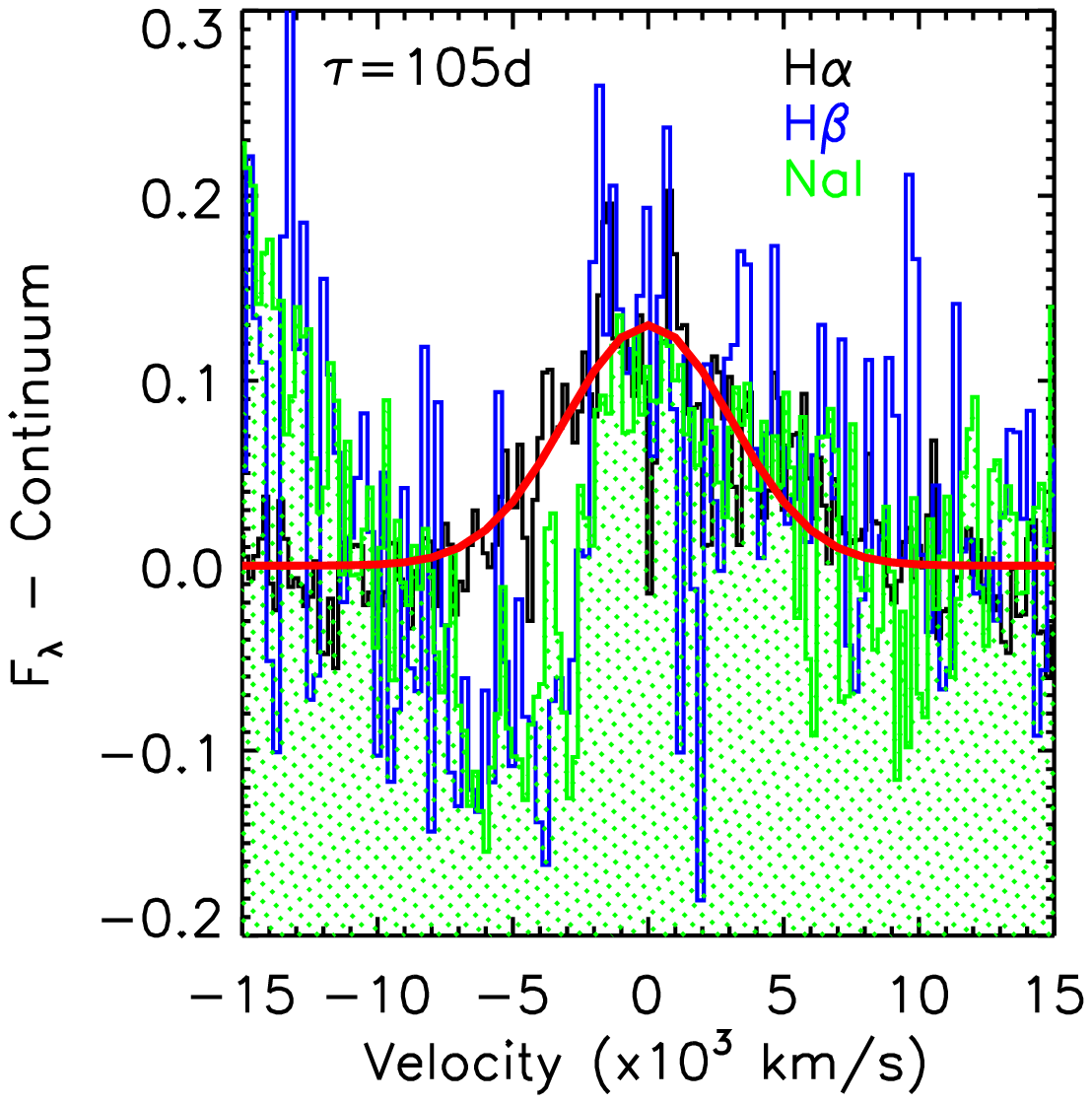}
\caption{Left: broad He II $\lambda$4686 in the 2008 May 1 and 8 spectra, with the continuum subtracted, and the best-fit Gaussian with FWHM = (5.5 $\pm 0.5) \times 10^{3}$ km s$^{-1}$ (red line).  Right: broad H$\alpha$, H$\beta$, and Na I $\lambda$5892 emission lines with the continuum subtracted.  The H$\beta$ and Na I lines have been scaled by a factor of 3 to match the flux of H$\alpha$.  The single Gaussian fit to the broad H$\alpha$ line with FWHM = (1.01 $\pm 0.06) \times 10^{4}$ km s$^{-1}$ is shown with a red line, and the area under the Na I line is shaded in green.  The H$\beta$ and Na I lines have a strong blueshifted P Cygni absorption feature with a minimum at $-7 \times 10^{3}$ km s$^{-1}$ that is not seen in the profile of the H$\alpha$ line.
\label{fig_vel}
}
\end{figure*}

Strong Balmer lines and Na I $\lambda 5892$ did not appear until after more than $20$ rest-frame days after the explosion.  The broad H$\alpha$ line in the 2008 August 1 spectrum can be fitted with a single Gaussian with an FWHM = (1.02$ \pm 0.06) \times 10^{4}$ km s$^{-1}$.  The spikes near the peak of H$\alpha$ are residuals from a poor sky subtraction.  The broad H$\alpha$ flux is ($3.7 \pm 0.4$) $\times 10^{-15}$ erg s$^{-1}$ cm$^{-2}$, corresponding to a luminosity of $(3.7 \pm 0.4) \times 10^{41}$ erg s$^{-1}$.  The Na I $\lambda 5892$ and H$\beta$ lines have P Cygni line profiles with a blue-shifted absorption component with a minimum at a velocity of $(-7.4 \pm 0.2) \times 10^{3}$ km s$^{-1}$ and $(-7 \pm 1) \times 10^{3}$ km s$^{-1}$ respectively, and a blue absorption wing that extends out to $\sim 1 \times 10^{4}$ km s$^{-1}$.  Figure \ref{fig_vel} shows the broad emission lines, and their best fitting Gaussian profiles.  The lack of a strong P Cygni absorption component in the H$\alpha$ line and P Cygni profiles in the H$\beta$ and Na I lines is consistent with the bright Type II-L SNe 1979C and 1980K (Panagia \etal 1980; Branch \etal 1981; Uomoto \& Kirshner 1986).    The width of the H$\alpha$ line is a measure of the expansion speed of the SN ejecta, while the absorption component of the P Cygni profile traces the outflow of the absorbing material in the line of sight.  Both components indicate an SN blast wave velocity of $\sim$ 10,000 km s$^{-1}$.  

The error in the redshift from the Gaussian fit to the He II $\lambda 4686$ line does not reflect systematic errors that may be introduced if the He II line is blended or is not symmetric.  The best-fitting SN II template in the Supernova Identification Code (SNID) database (Blondin \& Tonry 2007) to the spectrum 105 days after explosion is SN II-L 1990K on day +53 (Cappellaro \etal 1995) at $z=0.202 \pm 0.006$.  A fit to all Type II-L SNe in the database gives the same best-fit result.  If we restrict the fit to SN templates 1979C and 1980K (as is done by Miller \etal 2008c), a worse fit is found, with $z=0.213 \pm 0.011$.  The redshift measured from the best template fits is consistent with the redshift determined from the He II line, and so we use this redshift but change the error bars to reflect the uncertainties in the template fit, $z=0.205^{+.003}_{-.009}$. 

\newpage

\begin{figure*}
\epsscale{0.4}
\begin{center}
\epsfig{file=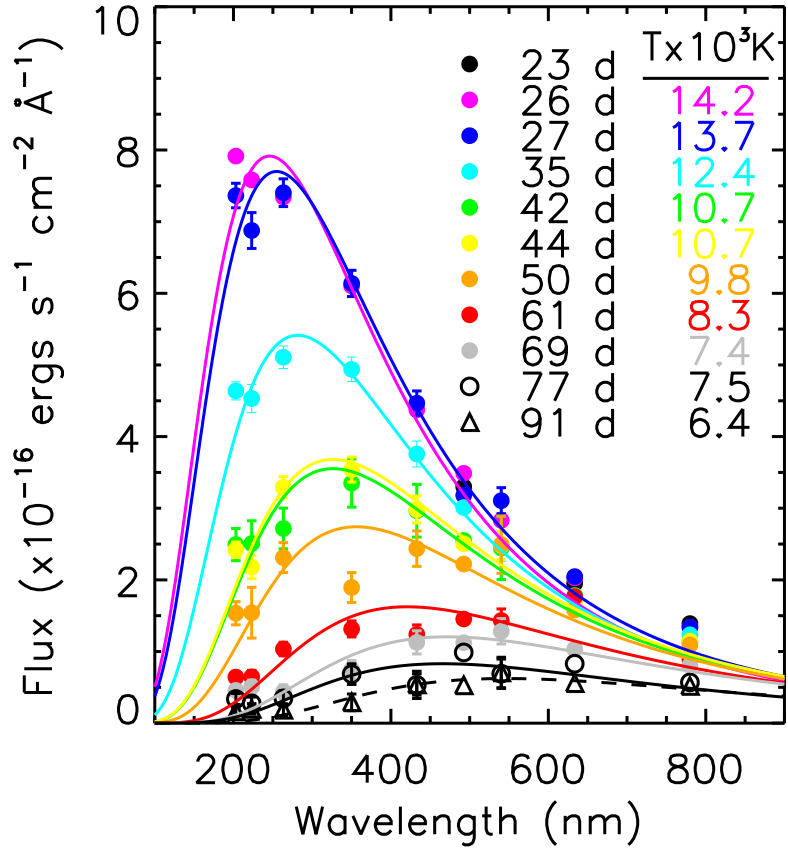}
\caption{SED of SN 2008es over time in observed days since the time of explosion measured by the \textsl{Swift} UVOT telescope in the $uvw2$, $uvm2$, $uvw1$, $u$, $B$, and $V$ bands, and the Palomar 60 inch telescope in the $g$, $r$, and $i'$ bands.  \textsl{Swift} UVOT telescope magnitudes are converted to flux densities using the count rate to flux conversion from the Pickles standard stars (Poole \etal 2008).  Temperature fits are shown for a rest-frame blackbody at $z=0.205$.
\label{fig_sed}
}
\end{center}
\end{figure*}

\begin{figure}
\plotone{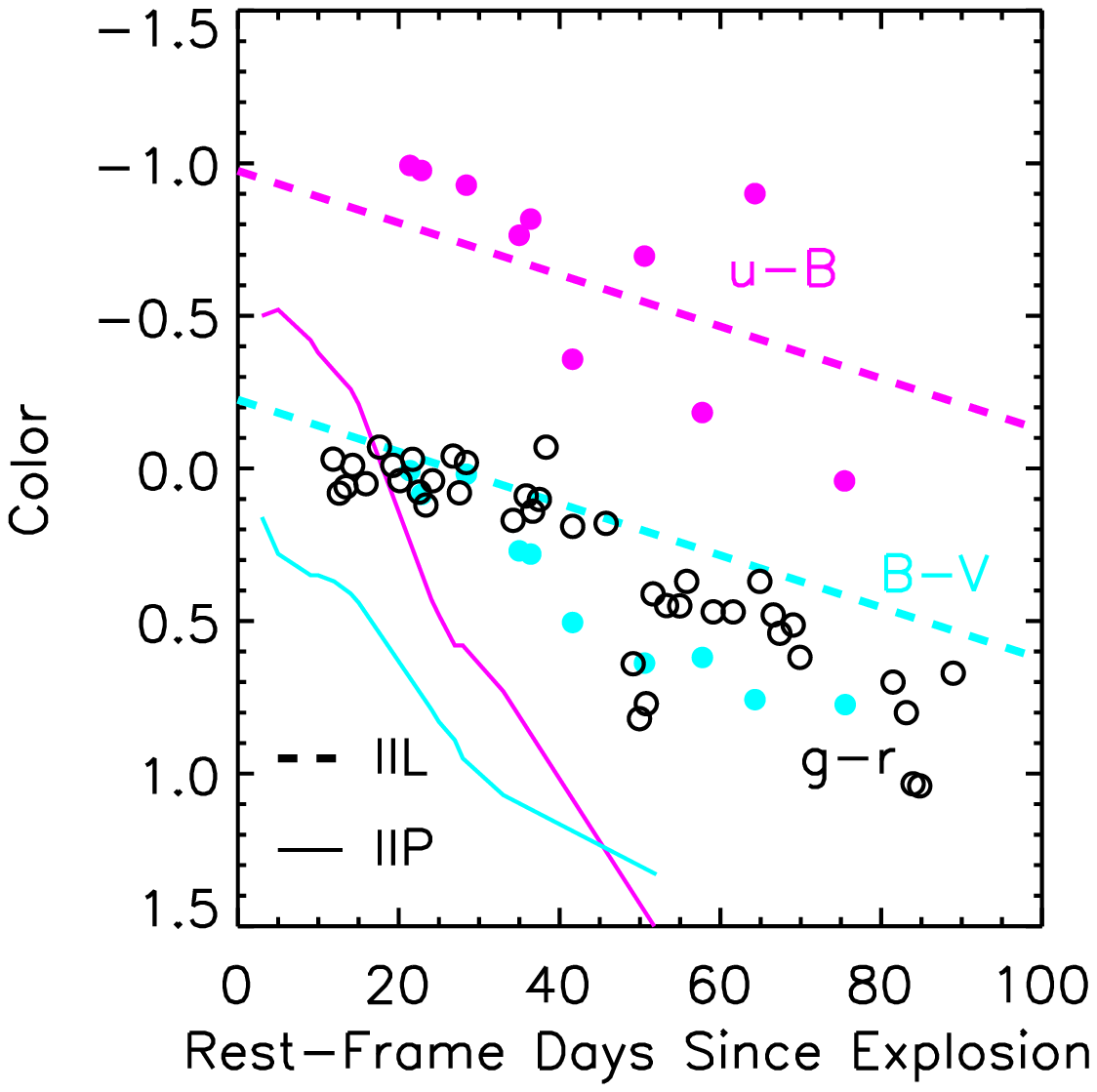}
\caption{\textsl{Swift} $u-B$ and $B-V$ , and P60 $g-r$ colors of SN 2008es over time in the SN rest frame in comparison to Type II-L SN 1979C and Type II-P SN 2006bp.
\label{fig_colors}
}
\end{figure}

\subsection{Host Galaxy} \label{sect_image}
No host galaxy is detected in the pre-SN Sloan Digital Sky Survey (SDSS) Data Release 6 (DR6) images (Adelman-McCarthy \etal 2008), down to a limiting magnitude of $r \sim 23$.  This constrains the host galaxy to be a dwarf galaxy with $M_{r} \apgt -17$ at $z=0.205$.  The closest galaxy detected by SDSS is located 8$\farcs$3 North East of the SN, with $r=23.1 \pm 0.4$ and $g-r = -0.1\pm0.5$\,mag.  At a redshift of $z=0.205$, this angular separation corresponds to a projected separation of $\sim 40$ kpc, which is much larger than that would be expected if the galaxy were the SN host.  During our 2008 August 1 spectrum, we placed the slit at a position angle so that this galaxy would fall in the slit, but we did not detect any emission lines from the galaxy that could be used to determine its redshift.  The accuracy of the SN redshift could be greatly improved with future spectroscopic observations at the position of the SN, after the SN has faded significantly, when emission lines from the host galaxy that are now outshined by the SN may be detectable.

\subsection{Swift Observations}\label{sect_swift}
We requested Target of Opportunity (ToO) observations with the Ultraviolet/Optical Telescope (UVOT; Roming \etal 2005) and X-Ray Telescope (XRT; Burrows \etal 2005) on board the \textsl{Swift} observatory (Gehrels \etal 2004) on 2008 May 14, with continued monitoring through 2008 September 2.  The spectral energy distribution (SED) of the source on 2008 May 14 was well described by a blackbody with $T \sim 10^{4}$ K for $z \sim 1$.  The corresponding bolometric luminosity of $\sim 10^{46}$ erg s$^{-1}$, in combination with the lack of X-ray emission indicative of persistent quasar activity, suggested that the source may have been a flare from the accretion of a tidally disrupted star around a central black hole of $10^{8}$ \msun (Gezari \& Halpern 2008).  This interpretation was soon ruled out, however, by the rapidly cooling temperature of the emission seen in the later \textsl{Swift} observations, and the subsequent appearance of SN-like broad features as the spectra evolved (Miller \etal 2008a; Gezari \etal 2008b). 

No X-ray source was detected in any of the individual XRT epochs.  The count rate in a 23$\arcsec$ extraction radius for the accumulated exposure time of 50.27 ks is $4.8 \times 10^{-4}$ counts s$^{-1}$, and places a 3$\sigma$ upper limit to the 0.3-10 keV flux, using the standard energy conversion factor, of $2.4 \times 10^{-14}$ erg s$^{-1}$ cm$^{-2}$ s$^{-1}$, which corresponds to a luminosity at $z=0.205$ of $L_{X} < 2.4 \times 10^{42}$ erg s$^{-1}$, using a luminosity distance for $z=0.205$ of $d_{L}=1008$ Mpc ($H_{0}$= 70 km s$^{-1}$ Mpc$^{-1}$, $\Omega_{m}$=0.30 and $\Omega_{\Lambda}$=0.70). 

The UVOT magnitudes were measured with a 3$\arcsec$ aperture, and an aperture correction based on an average point-spread function (PSF).  We convert the magnitudes into flux densities using the count rate to flux conversions determined for the Pickle standard stars (Poole \etal 2008).
The UVOT $uvw2 (\lambda_{eff}=203$nm), $uvm2 (\lambda_{eff}=223$nm), $uvw1 (\lambda_{eff}=263$nm), $u (\lambda_{eff}=350$nm), $b (\lambda_{eff}=433$nm), and $v (\lambda_{eff}=540$nm) band magnitudes are plotted in Figure \ref{fig_lc} and listed in Table \ref{tab_3}.  In Figure \ref{fig_sed} we show the SED of the SN over time from 200 to 800 nm measured by UVOT in combination with optical data points from the P60 data closest in time to the UVOT observations.  The UVOT $b$ and $v$ bands are interchangeable with the Johnson $B$ and $V$ system, but the $u$ band has a bluer response than the Johnson $U$ band, and we make this distinction throughout the paper.

We fit the SEDs over time with a blackbody at $z=0.205$, and fit temperatures starting $t = 26$ days after explosion with 14,000 K, cooling down to 6400 K at $t = 91$ days, with the corresponding blackbody radii expanding from $3 \times 10^{15}$ cm to $5 \times 10^{15}$ cm.  These radii are $\apgt 50$ times larger than a standard red supergiant photosphere, and mostly likely correspond to the radius of the expanding SN blast wave.  We see evidence of an excess above of a pure blackbody in the $uvw2$ band at early times ($26-27$ observed days after explosion), that may be the effect of strong FeIV and FeIII features in that region, seen in the spectra of SN II-P 2006bp at similar photospheric temperatures (Dessart \etal 2008).  

\section{Results}\label{sec_disc}

\subsection{Comparison with Type II SNe}

The high signal-to-noise (S/N) MDM 1.3m $V$-band observations, with $\Delta$ mag $\sim 0.01$, measure a linear decay from 9 to 53 d after the peak of 2.95 $\pm 0.02$ mag (100 days)$^{-1}$, or 3.55 mag (100 days)$^{-1}$ in the SN rest frame.  This decay rate is just above the transition of the average decay-rates in the first 100 days after maximum from Type II-plateau (II-P) to Type II-linear (II-L) classes of SNe defined by \citet{1994A&A...282..731P} to be 3.5 mag (100 days)$^{-1}$ in the $B$ band (at $z\sim0.2$ the $V$ band is close to the rest-frame $B$ band).  This classification is nonexclusive, however, since \citet{1994A&A...282..731P} point out that there is a continuum in rates between the plateau and linear subclasses.  Type II-L SNe are on average brighter than Type II-P SNe, and II-L SNe appear to have a bimodal distribution of absolute peak magnitudes, with a subclass of bright II-Ls with $<{M}_{B}> = -19.27$ \citep{2002AJ....123..745R}.  Yet, even the bright II-L SNe are still 3 mag fainter than the peak magnitude of SNe 2008es. 

The \textsl{Swift} and MDM $V$-band observations show evidence of a flattening of the optical light curve $\sim 65$ rest-frame days after maximum, with a slope that is consistent with the $^{56}$Co decay rate of 0.98 mag (100 days)$^{-1}$ (shown with a dotted line in Figure \ref{fig_lc}, stretched in time to match the time dilation of SN 2008es).  The estimated luminosity at the time of the MDM $V$-band data point at 104 days after explosion in the SN rest frame, is $\approx 1 \times 10^{42}$ erg s$^{-1}$ (a factor of $\sim$ 3 below the luminosity on day 75 in the SN rest frame), and using the expression, $L = 1.42 \times 10^{43} e^{-t_{d}/111} M_{Ni}/$\msun~erg s$^{-1}$ \citep{1984ApJ...280..282S}, implies an upper limit to the initial $^{56}$Ni mass of $\aplt 0.2$ \msun, which is close to the upper end of the range masses measured for Type II SNe (e.g., 0.3 \msun for SN 1992am; Schmidt \etal 1994).  

In Figure \ref{fig_colors}, we compare the $u-B$, $B-V$, and $g-r$ colors of SN 2008es over time to the evolution of the $U-B$ and $B-V$ colors of SN II-L 1979C which have a linear slope of $\sim 0.85$ mag (100 days)$^{-1}$ (Panagia \etal 1980).  The $K$-correction at the redshift of SN 2008es makes the $g-r$ color close to rest-frame $B-V$.  We also show the $u-B$ and $B-V$ light curves of SN II-P 2006bp measured by \textsl{Swift} (Immler \etal 2007).  All curves are shown in days since explosion in the SN rest frame.  The light curve of SN 2008es has a much slower evolution of the $B-V$ color than seen in prototypical Type II-P SN 2006bp, and is in closer agreement with the colors and linear slopes of prototypical bright Type II-L SN 1979C.  

\begin{figure}
\plotone{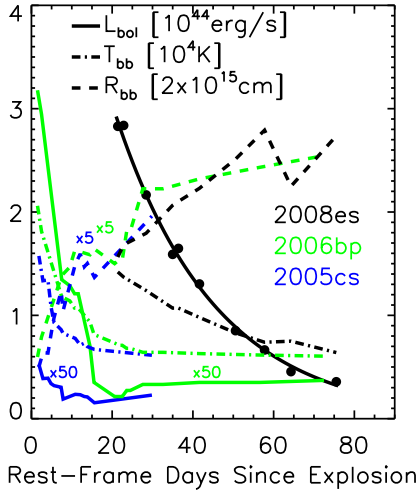}
\caption{Blackbody temperature, radius, and bolometric luminosity over time in the SN rest frame for SN 2008es, and Type II-P SNe 2005cs and 2006bp.  The luminosity and radius of SNe 2005cs and 2006bp have been scaled by a factor of 50 and 5, respectively, for comparison with SN 2008es.
\label{fig_lbol}
}
\end{figure}

We estimate the bolometric luminosity of SN 2008es over time (as shown in Figure \ref{fig_lbol}) using the blackbody curve fits shown in Figure \ref{fig_sed}.  The luminosity during the optical peak, 23 days after the explosion in the SN rest frame, is $2.8 \times 10^{44}$ erg s$^{-1}$, and decays exponentially as $L_{bol}(t) = L_{bol}(t=0) e^{-\alpha t}$ erg s$^{-1}$, where $L_{bol}(t=0) = 6.5 \times 10^{44}$ erg s$^{-1}$, $\alpha = 0.040 \pm 0.004$, and $t$ is the rest-frame days since explosion.  We estimate the total energy radiated by integrating under the curve from day 21 to 75 in the SN rest frame to get $E_{\rm tot}=5.6 \times 10^{50}$ erg, a value comparable to the radiated energy of the extreme SN 2006gy (Ofek \etal 2007; Smith \etal 2007) which emitted $\sim 1 \times 10^{51}$ erg s$^{-1}$ in the first two months.  These radiated energies, so close to the total kinetic energy produced in a core-collapse explosion (typically in the range of $0.5-2.0 \times 10^{51}$ erg), require an efficient conversion of the kinetic energy of the SN ejecta into radiation.

Figure \ref{fig_lbol} shows the evolution of the bolometric luminosity of SN 2008es in comparison to two Type II-P SNe with detailed non-LTE modeling of their photometry and spectra by \citet{2008ApJ...675..644D}: SNe 2005cs and 2006bp.  The luminosity curves have been scaled by a factor of 50 for comparison with SN 2008es.  The behavior of SN 2008es is markedly different than the SNe II-P, with a factor of 50 greater luminosity at 23 days after explosion compared to the peak luminosity of SN 2006bp that occurs a couple of days after explosion, and with a much slower decline in luminosity over 75 days in comparison to $\sim 15$ days in SNe 2005cs and 2006bp. 
Figure \ref{fig_lbol} also shows $T_{bb}$ and $R_{bb}$ for SN 2008es in comparison to the Type II-P SNe 2005cs and 2006bp, with the radii of the Type II-P SNe scaled up by a factor of 5 for comparison.  In contrast to SN 2008es which has $T=14,000$ K at 23 days after explosion, the photospheric temperatures of the Type II-P SNe have already cooled to 6000 K by day 20, and their photospheric radii are $\sim 5$ times smaller.   It is clear that there is a different mechanism powering the long-lived, and extremely luminous emission from SN 2008es.  

\subsection{Interaction with CSM}

Given the estimate of the expansion speed, $v_{s}$, from the broad H$\alpha$ line of 10,000 km s$^{-1}$, the radius of the ejecta at the time of the peak of the optical light curve,  $t_{peak} = 23$ days after explosion in the SN rest frame, is expected to be $R_{ph}=v_{s}t_{peak} = 2 \times 10^{15}$ cm, which is in good agreement with the radius of the blackbody fit at that time ($R_{bb} \sim 3 \times 10^{15}$ cm).  If the luminosity during the optical maximum is powered by interaction with the CSM, this requires a density of order $n \sim L/(2\pi R^{2}v_{s}^{3}m) \sim 2 \times 10^{9}$ cm$^{-3}$, where $m$ is the mean mass per particle (2.1 $\times 10^{-24}$ g for H+He) and we use $R_{\rm ph} = 3 \times 10^{15}$ cm.  This implies an enclosed mass within $R_{\rm ph}$ of $\sim 0.2$ \msun, and corresponds to a mass-loss rate of $\dot{M} \sim M/(tv_{s}/v_{w}) \sim 2.5 \times 10^{-3} (v_{w}$/10 km s$^{-1}$) \msun yr$^{-1}$, where $v_{w}$ is the velocity of the stellar wind ($\sim 10$ km s$^{-1}$ for a red supergiant progenitor).

We can place an upper limit on the progenitor's mass-loss rate if we assume that X-ray emission is produced by the interaction of the SN shock with the ambient CSM.  In the forward shock, a H+He plasma will be heated to $T = 1.36 \times 10^{9} [(n-3)/(n-2)]^{2}v_{s4}^{2}$ K (Chevalier \& Fransson 2003), where $n$ is the ejecta density parameter (in the range of $7-12$) and $v_{s4}$ is the shock velocity in $10^{4}$ km s$^{-1}$, and will produce an X-ray luminosity of $L_{X}=3.3\times10^{60}\Lambda(T)(\dot{M_{-6}}/v_{w1})^{2}(v_{s4}t_{d})^{-1}$ erg s$^{-1}$ (Immler \etal 2002), where $v_{w1}$ is the stellar wind speed in units of 10 km s$^{-1}$, $t_{d}$ is the time since explosion in days, $\dot{M}_{-6}$ is the progenitor's mass-loss rate in units of 10$^{-6}$ \msun yr$^{-1}$, and $\Lambda(T) = 2.4 \times 10^{-23} g_{\rm ff} T_{8}^{0.5}$ is the cooling function for an optically thin heated plasma, where $g_{\rm ff}$ is the free-free Gaunt factor, and $T_{8}$ is the temperature in units of $10^{8}$ K. 
Using the width of the broad H$\alpha$ line to estimate $v_{s4} \sim 1$, and using $t_{d}=23$, the upper limit on $L_{X}(0.3-10)$ keV implies that $\dot{M} \aplt 5 \times 10^{-4} v_{w1}$ \msun yr$^{-1}$.  
This upper limit to the mass-loss rate is still a factor of $5$ below what is required to power the observed peak bolometric luminosity purely by interaction with the CSM. 

\subsection{Comparison with Models}

The $B$-band light curve and $B-V$ colors of bright SN II-L 1979C were successfully modeled by Blinnikov \& Bartunov (1993) with a red supergiant progenitor with an extended structure ($R \sim 6000$ \rsun), a small hydrogen envelope mass ($\sim 1-2$ \msun), and a dense superwind with a mass-loss rate of $\dot{M} \approx 10^{-4}$ \msun yr$^{-1}$.  In this model, the large peak luminosity of SN 1979C ($M_{B}=-19.6$) is produced by the dense stellar wind that reprocesses the UV photons produced at shock breakout, and can be boosted up to $M_{B} \sim -22$ by increasing the mass-loss rate to $\dot{M} \approx 10^{-3}$ \msun yr$^{-1}$.  
Their model predicts a steep drop in luminosity 50 days after explosion, when the recombination (in the diffusion regime) reaches the silicon core.  After this point, the light curve has an extended tail powered by radioactive decay and/or emission from the shock propagating into the CSM.  These predictions are consistent with the sharp drop and subsequent $^{56}$Co decay slope tentatively detected in the $V$-band light curve of SN 2008es. The Blinnikov \& Bartunov (1993) model also explains the absence of the absorption component in H$\alpha$, which is due to the fact that at late stages the temperature inside the dense envelope is lower than the shock wave temperature.

Young \etal (2005) find these values for the radius and superwind unrealistic, and prefer a model in which the bright luminosity at the early phase of the SN light curve is the result of an optical afterglow from a GRB jet interacting with the progenitor's hydrogen envelope.  The main problem with this model is that the optical afterglow emission produced from the external shocks created by the interaction of the relativistic jet with the H envelope will emit nonthermal synchotron radiation, which is in direct contradiction with the cooling blackbody SED of SN 2008es. 

Another possibility, favored by Miller \etal (2008c), is that the luminosity is powered by the interaction of the SN blast wave with a massive circumstellar shell that was ejected in an episodic eruption (Smith \& McCray 2007).  Similarly, the luminosity could be powered by an interaction between such shells (Woosley \etal 2007).  Both of these models produce their luminosity via the dense shell that decelerates the blast wave, and converts the bulk kinetic energy into radiation through shocks.  The intermediate-width (a few thousand km s$^{-1}$) P Cygni lines produced from the dense postshock gas, and the slow escape of radiation due to photon diffusion of the thermal energy through the opaque shell in these models with massive shells are not consistent with the broad, symmetric H$\alpha$ line emission, and relatively fast rise time of SN 2008es.  

If the bolometric luminosity of SN 2008es was powered solely by thermalization of gamma rays from radioactive cobalt decay, then the luminosity at $t_{d}=23$ would require 25 \msun\ of $^{56}$Ni to be synthesized in the explosion.  This is 3 orders of magnitude larger than that typically produced in Type II SNe, and much larger than the ejecta mass which is expected to be less than 5\,\msun\ to avoid a plateau from H recombination in the optical light curve.  Although it is possible for such a large amount of $^{56}$Ni to be produced by an extremely massive progenitor star ($> 100$ \msun) that underwent a pair-instability supernova (Ober \etal 1983), we already find evidence for a transition to a radioactive heating-dominated light curve at late times, which suggests that the light curve was not dominated by radioactive decay near the peak. Furthermore, the calculations of Scannapieco \etal (2005) predict a long, slow rise to maximum, and slow expansion speeds of $\sim 5000$ km s$^{-1}$, which is neither consistent with the 23 days rise to maximum nor the 10,000 km s$^{-1}$ expansion speed seen in SN 2008es.  

Given the contradictions with the more exotic scenarios discussed above, the extreme luminosity of Type II-L SN 2008es seems most likely to be on the bright end of a range of luminosities and light curves possible with variations of progenitor star H envelope and stellar wind parameters presented in the models by Blinnikov \& Bartunov (1993).  However, even the largest red supergiant stars have radii \aplt $1500$ \rsun (Levesque \etal 2005), thus a progenitor star with the extreme extended structure assumed by Blinnikov \& Bartunov (1993) is problematic.  Type II-L SN light curves with peak magnitudes of $M_{B} \sim -19$ were reproduced by Swartz \etal (1991) for core-collapse models with more realistic envelope radii (1200 \rsun).   More detailed modeling is needed to determine if the required parameters for SN 2008es are consistent with stellar evolution.

\begin{figure*}
\epsscale{0.25}
\begin{center}
\epsfig{file=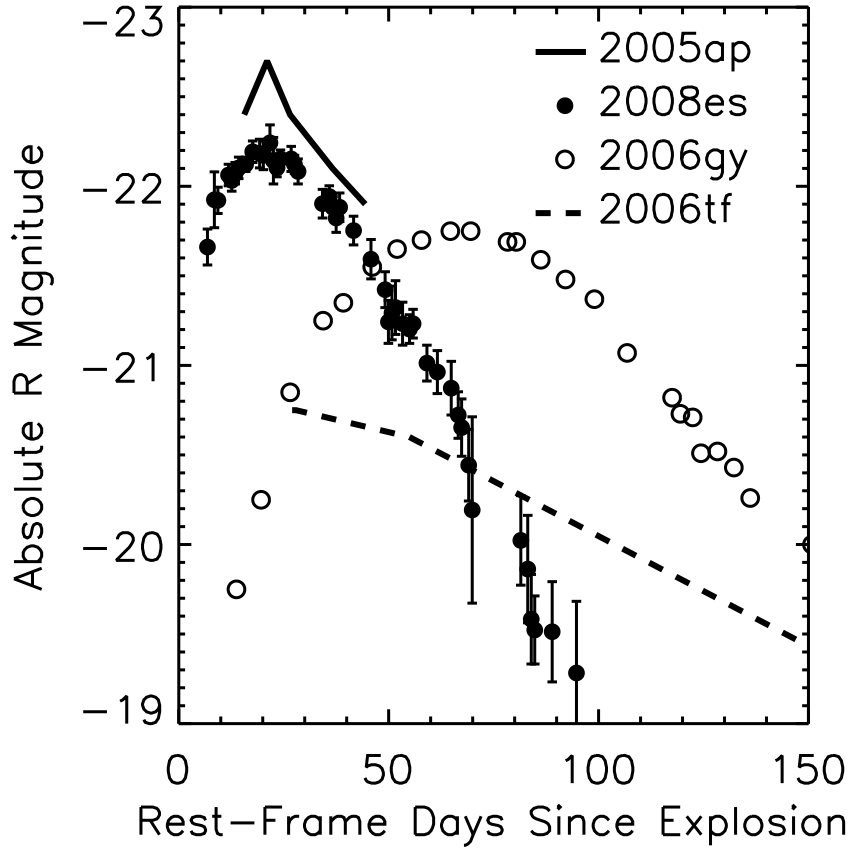}
\caption{
Comparison of absolute $R$ magnitudes of the four overluminous SNe discovered by the ROTSE-IIIb telescope.  The light curve of SN 2008es is constructed from the unfiltered ROTSE-IIIb observations and the P60 $r$-band observations.  The light curves are shown corrected for time dilation.  The day of explosion for SNe 2005ap is estimated by Quimby \etal (2007a) to be $\sim 3$ weeks before the peak in the rest frame of the SN at $z=0.2832$.  The rise time of SN 2006tf is not well constrained, so for comparison, we assume that the peak occurs at the same time after explosion as SN 2008es. 
\label{fig_sne}
}
\end{center}
\end{figure*}

\section{Conclusion}\label{sec_conc}

The overluminous SN 2008es ($M_{V}=-22.2)$ is the second most luminous SN observed, and in the same league as the extreme SNe 2005ap, 2006gy, and 2006tf, which had peak visual magnitudes of $-22.7$, $-22.0$, and $-20.7$, respectively.  In Figure \ref{fig_sne}, we compare the absolute $R$-band light curves of the 4 SNe (all discovered by the ROTSE-IIIb telescope) in rest-frame days since the time of explosion.  SN 2006gy and 2006tf are classified spectroscopically as Type IIn SNe because of their narrow peaked P Cygni Balmer lines (Ofek \etal 2007; Smith \etal 2008), although with deviations from the smooth profiles typical of SNe IIn in SN 2006gy (Smith \etal 2007).  The time of explosion for SN 2005ap is estimated by Quimby \etal (2007a) to be $\sim 3$ weeks before maximum.  The time of explosion for SN 2006tf is poorly constrained (Smith \etal 2008), so for comparison we assume that it takes the same time to reach the peak as SN 2008es.   

Of the SNe in the same luminosity class, the light curve of SN 2008es is most like SN 2005ap, the most luminous SN ever observed \citep{2007ApJ...668L..99Q}. 
Interestingly, SN 2005ap also has a dwarf galaxy host, with $M_{R} = -16.8$.    However, SN 2005ap does differ from SN 2008es in the fact that it demonstrated absorption features at early times, as well as P Cygni absorption in its broad H$\alpha$ profile.  This is not consistent with the spectra of bright II-L SNe which have decreasing P Cygni absorption with increasing luminosity (Patat \etal 1994).  The extended envelope/wind model is problematic for SN 2005ap, since it would be hard to produce early absorption features without corresponding emission.
The extreme luminosity and linear decay of SN 2005ap were attributed instead to a collision of the SN ejecta with a dense circumstellar shell, a GRB explosion buried in a H envelope, or a pair-instability eruption powered by radioactive decay \citep{2007ApJ...668L..99Q}.

The peak luminosity, linear decay, and spectroscopic features of SN 2008es place it in a subclass of ultra-bright Type II-L SNe. 
The light curve and SED of SN 2008es, measured in detail from the UV through the optical, point toward a core-collapse explosion of a nonstandard progenitor star with a super wind and extended envelope.  Wide-field optical synoptic surveys such as the ROTSE-III Supernova Verification Project, Pan-STARRS, and LSST will continue to explore the large parameter space of core-collapse explosions, and increase our understanding of the effects of variations in progenitor star properties and environments.

{\bf Note Added in Proof:} More recent $V$-band observations from \textsl{Swift}, as well as $V, R, I$ measurements from the MDM 2.4m, reveal that the slope of the light curve is steeper than the decay rate of $^{56}$Co up to $\sim$145 rest-frame days after maximum.  Therefore, radioactive decay must not yet be the dominant source of the luminosity from the SN, and the inferred upper limit on the initial $^{56}$Ni mass is securely within the range of normal Type II SNe (Hamuy 2003).

\acknowledgements

We thank the \textsl{Swift} PI Neil Gehrels for approving our various  ToO monitoring requests for SN 2008es, the \textsl{Swift} Science Operating Center team for performing the observations, Don Schneider for approving our HET ToO request, and the HET resident astronomers for performing our HET spectroscopic follow-up observations. The Hobby-Eberly Telescope (HET) is a joint project of the University of Texas at Austin, the Pennsylvania State University (PSU), Stanford University, Ludwig-Maximilians-Universit\"at M\"unchen, and Georg-August-Universit\"at G\"ottingen. The HET is named in honor of its principal benefactors, William P. Hobby and Robert E. Eberly.  \textsl{Swift} is supported at PSU by NASA contract NAS5-00136.  ROTSE-III has been supported by NASA grant NNG-04WC41G, the Australian Research Council, the University of New South Wales, the University of Texas, and the University of Michigan.  P60 operations are funded in part by NASA through the \textit{Swift} Guest Investigator Program (grant NNG06GH61G).  F.Y.~is supported by NASA grants NNX-07AF02G and NNX-08AN25G.
J.C.W.~is supported in part by NSF grant AST 0707769.

\clearpage
\LongTables
\begin{deluxetable}{lccc}
\tablewidth{0pt}
\tablecaption{Ground-based Optical Photometry \label{tab_1}}
\tablehead{
\colhead{Telescope/Band} & \colhead{MJD\tablenotemark{a}} & \colhead{Magnitude\tablenotemark{b}} & \colhead{Error}
}
\startdata
ROTSE-IIIb & 54555.1 & (19.48) & \\
 & 54556.2 & (19.51) & \\
 & 54559.1 & (18.32) & \\
 & 54562.1 & (19.21) & \\
 & 54567.4 & (18.93) & \\
 & 54569.2 & (18.69) & \\
 & 54571.2 & (18.25) & \\
 & 54573.2 & (18.46) & \\
 & 54575.2 & (18.54) & \\
 & 54579.2 & (18.27) & \\
 & 54580.2 & (18.34) & \\
 & 54582.2 & 18.36 & 0.10 \\
 & 54584.2 & 18.09 & 0.16 \\
 & 54585.2 & 18.10 & 0.07 \\
 & 54591.2 & 18.02 & 0.18 \\
 & 54592.2 & 17.96 & 0.22 \\
 & 54594.2 & 17.78 & 0.10 \\
 & 54596.2 & 17.79 & 0.09 \\
 & 54597.1 & 17.66 & 0.21 \\
 & 54604.2 & 17.95 & 0.15 \\
 & 54605.2 & 17.91 & 0.18 \\
 & 54606.2 & 17.98 & 0.08 \\
 & 54608.3 & 18.05 & 0.33 \\
 & 54610.2 & 17.87 & 0.05 \\
 & 54618.2 & 17.98 & 0.06 \\
 & 54619.2 & 18.14 & 0.09 \\
 & 54620.2 & 18.09 & 0.09 \\
 & 54622.2 & 18.21 & 0.15 \\
 & 54624.2 & 18.50 & 0.16 \\
 & 54626.2 & 18.14 & 0.14 \\
 & 54627.2 & 18.60 & 0.16 \\
P60/$g$ & 54588.3 & 17.95 & 0.06 \\
 & 54589.2 & 17.98 & 0.06 \\
 & 54590.2 & 17.94 & 0.06 \\
 & 54591.2 & 17.91 & 0.06 \\
 & 54593.2 & 17.89 & 0.04 \\
 & 54595.2 & 17.82 & 0.06 \\
 & 54597.2 & 17.83 & 0.08 \\
 & 54598.3 & 17.84 & 0.08 \\
 & 54600.2 & 17.77 & 0.10 \\
 & 54601.2 & 17.87 & 0.13 \\
 & 54602.2 & 17.91 & 0.05 \\
 & 54603.2 & 17.86 & 0.05 \\
 & 54606.3 & 17.86 & 0.07 \\
 & 54607.2 & 17.89 & 0.05 \\
 & 54608.2 & 17.93 & 0.07 \\
 & 54615.2 & 18.11 & 0.08 \\
 & 54617.2 & 18.07 & 0.06 \\
 & 54618.2 & 18.13 & 0.06 \\
 & 54619.2 & 18.19 & 0.08 \\
 & 54620.2 & 18.13 & 0.08 \\
 & 54624.2 & 18.26 & 0.08 \\
 & 54629.2 & 18.42 & 0.11 \\
 & 54633.2 & 18.59 & 0.10 \\
 & 54634.2 & 18.77 & 0.12 \\
 & 54635.2 & 18.72 & 0.15 \\
 & 54636.2 & 18.69 & 0.15 \\
 & 54638.2 & 18.78 & 0.12 \\
 & 54640.2 & 18.81 & 0.08 \\
 & 54641.2 & 18.78 & 0.08 \\
 & 54645.2 & 19.00 & 0.10 \\
 & 54648.2 & 19.05 & 0.12 \\
 & 54652.2 & 19.14 & 0.15 \\
 & 54654.2 & 19.29 & 0.13 \\
 & 54655.2 & 19.36 & 0.16 \\
 & 54657.2 & 19.57 & 0.20 \\
 & 54658.2 & 19.82 & 0.52 \\
 & 54672.2 & 19.99 & 0.25 \\
 & 54674.2 & 20.15 & 0.30 \\
 & 54675.2 & 20.43 & 0.25 \\
 & 54676.2 & 20.49 & 0.19 \\
 & 54681.2 & 20.50 & 0.28 \\
 & 54688.2 & 20.73 & 0.40 \\
P60/$r$ & 54588.3 & 18.00 & 0.06 \\
 & 54590.2 & 17.90 & 0.06 \\
 & 54591.2 & 17.94 & 0.06 \\
 & 54593.2 & 17.86 & 0.06 \\
 & 54595.2 & 17.91 & 0.08 \\
 & 54597.2 & 17.86 & 0.08 \\
 & 54598.3 & 17.82 & 0.08 \\
 & 54600.2 & 17.82 & 0.10 \\
 & 54602.2 & 17.81 & 0.16 \\
 & 54603.2 & 17.84 & 0.08 \\
 & 54606.3 & 17.92 & 0.10 \\
 & 54607.2 & 17.83 & 0.08 \\
 & 54608.2 & 17.97 & 0.12 \\
 & 54615.2 & 17.96 & 0.08 \\
 & 54617.2 & 18.00 & 0.06 \\
 & 54618.2 & 18.01 & 0.06 \\
 & 54620.2 & 18.22 & 0.14 \\
 & 54624.2 & 18.09 & 0.08 \\
 & 54629.2 & 18.26 & 0.07 \\
 & 54633.2 & 18.28 & 0.08 \\
 & 54634.2 & 18.33 & 0.09 \\
 & 54635.2 & 18.30 & 0.08 \\
 & 54636.2 & 18.30 & 0.11 \\
 & 54638.3 & 18.35 & 0.10 \\
 & 54640.2 & 18.38 & 0.06 \\
 & 54641.3 & 18.43 & 0.06 \\
 & 54645.2 & 18.55 & 0.08 \\
 & 54648.2 & 18.60 & 0.08 \\
 & 54652.2 & 18.79 & 0.15 \\
 & 54654.2 & 18.83 & 0.08 \\
 & 54655.2 & 18.84 & 0.09 \\
 & 54658.2 & 19.22 & 0.30 \\
 & 54672.2 & 19.31 & 0.15 \\
 & 54674.2 & 19.37 & 0.15 \\
 & 54676.2 & 19.47 & 0.15 \\
 & 54681.2 & 19.85 & 0.20 \\
 & 54688.2 & 19.95 & 0.29 \\
 & 54696.2 & 20.23 & 0.41 \\
P60/$i$ & 54588.3 & 18.13 & 0.08 \\
 & 54590.2 & 17.95 & 0.06 \\
 & 54591.2 & 17.92 & 0.08 \\
 & 54593.2 & 17.81 & 0.05 \\
 & 54595.2 & 17.88 & 0.06 \\
 & 54597.2 & 17.77 & 0.06 \\
 & 54600.2 & 17.81 & 0.07 \\
 & 54603.2 & 17.80 & 0.08 \\
 & 54606.3 & 17.87 & 0.10 \\
 & 54607.2 & 17.82 & 0.09 \\
 & 54608.2 & 17.90 & 0.10 \\
 & 54615.3 & 17.98 & 0.08 \\
 & 54617.2 & 17.97 & 0.10 \\
 & 54618.2 & 17.98 & 0.09 \\
 & 54620.2 & 18.00 & 0.08 \\
 & 54624.2 & 18.03 & 0.08 \\
 & 54629.2 & 18.12 & 0.08 \\
 & 54633.2 & 18.15 & 0.09 \\
 & 54634.2 & 18.16 & 0.07 \\
 & 54635.2 & 18.26 & 0.08 \\
 & 54636.2 & 18.24 & 0.09 \\
 & 54638.3 & 18.31 & 0.10 \\
 & 54640.2 & 18.35 & 0.06 \\
 & 54641.3 & 18.35 & 0.07 \\
 & 54645.2 & 18.36 & 0.07 \\
 & 54645.2 & 18.36 & 0.07 \\
 & 54655.2 & 18.74 & 0.09 \\
 & 54657.2 & 18.63 & 0.30 \\
 & 54658.2 & 18.84 & 0.25 \\
 & 54676.2 & 19.18 & 0.14 \\
 & 54681.2 & 19.39 & 0.17 \\
 & 54688.2 & 19.64 & 0.20 \\
MDM/$V$ & 54611.2 & 17.80 & 0.01 \\
 & 54614.1 & 17.89 & 0.02 \\
 & 54617.1 & 17.89 & 0.01 \\
 & 54621.1 & 18.01 & 0.01 \\
 & 54625.2 & 18.14 & 0.01 \\
 & 54626.2 & 18.17 & 0.01 \\
 & 54627.2 & 18.19 & 0.01 \\
 & 54628.2 & 18.23 & 0.01 \\
 & 54629.2 & 18.23 & 0.01 \\
 & 54630.2 & 18.28 & 0.01 \\
 & 54631.2 & 18.32 & 0.03 \\
 & 54632.2 & 18.35 & 0.01 \\
 & 54633.2 & 18.38 & 0.01 \\
 & 54635.2 & 18.41 & 0.02 \\
 & 54637.2 & 18.48 & 0.01 \\
 & 54638.2 & 18.53 & 0.01 \\
 & 54639.2 & 18.56 & 0.01 \\
 & 54640.2 & 18.60 & 0.01 \\
 & 54648.2 & 18.84 & 0.02 \\
 & 54655.2 & 19.10 & 0.02 \\
 & 54699.1 & 20.70 & 0.05 \\
P200/$B$ & 54701.2 & 21.23 & 0.12 \\
 & 54701.2 & 21.52 & 0.16 \\
P200/$V$ & 54700.1 & 20.81 & 0.10 \\
 & 54700.2 & 20.82 & 0.06 \\
 & 54701.2 & 20.64 & 0.06 \\
 & 54701.2 & 20.82 & 0.06 \\
P200/$i$ & 54700.1 & 20.40 & 0.16 \\
 & 54700.1 & 20.17 & 0.10 \\
 & 54700.2 & 20.15 & 0.03 \\
 & 54701.1 & 20.06 & 0.07 \\
 & 54701.1 & 20.14 & 0.04 \\
\enddata
\tablenotetext{a}{Days in JD-2,400,000.5}
\tablenotetext{b}{ROTSE-IIIb upper limits shown in parentheses.  P60, MDM, and P200 magnitudes corrected for a Galactic extinction of E($B-V$)=0.012 mag (Schlegel \etal 1998).}
\end{deluxetable}

\clearpage
\begin{landscape}
\begin{deluxetable}{lccclc}
\tablewidth{0pt}
\tablecaption{Log of Spectroscopic Observations \label{tab_2}}
\tablehead{
\colhead{MJD\tablenotemark{a}} & \colhead{Telescope/Instrument}  & \colhead{Exposure Time (s)} & \colhead{Grism\tablenotemark{b}} & \colhead{Slit} & \colhead{Resolving Power\tablenotemark{b}} 
}
\startdata
54586.8 & HET/LRS &  2x600  &     300 l/mm      &     2\farcs0 & 300 \\
54587.8 & P200/DBSP & 2x1200  & 600/4000 (b), 158/7500 (r) & 1\farcs0 &  1800 (b), 700 (r)  \\
54593.8 &  HET/LRS & 2x450   &    300 l/mm       &    2\farcs0  &  300 \\
54618.2 & HET/LRS & 3x900  &  600 l/mm      &    2\farcs0 & 650 \\  
54632.2  & HET/LRS & 3x900  &     600 l/mm    &       2\farcs0 & 650 \\
54678.7  & P200/DBSP & 2x1500 & 600/4000 (b), 158/7500 (r)  &  1\farcs0 & 1800 (b), 700 (r) \\
\enddata
\tablenotetext{a}{Days in JD-2,400,000.5}
\tablenotetext{b}{Values for the blue and red side of P200/DBSP are labeled (b) and (r).}
\end{deluxetable}

\clearpage
\end{landscape}
\begin{deluxetable}{lccc}
\tablewidth{0pt}
\tablecaption{\textsl{Swift} UVOT Photometry \label{tab_3}}
\tablehead{
\colhead{Band} & \colhead{MJD\tablenotemark{a}} & \colhead{Magnitude\tablenotemark{b}} & \colhead{Error}
}
\startdata
$uvw2$ & 54599.8 & 17.05 & 0.03 \\
 & 54601.5 & 17.13 & 0.03 \\
 & 54608.2 & 17.63 & 0.03 \\
 & 54616.1 & 18.30 & 0.09 \\
 & 54617.9 & 18.34 & 0.05 \\
 & 54624.1 & 18.83 & 0.11 \\
 & 54634.9 & 19.77 & 0.10 \\
 & 54643.6 & 20.15 & 0.14 \\
 & 54651.5 & 20.48 & 0.20 \\
 & 54664.9 & 21.26 & 0.29 \\
$uvm2$ & 54599.8 & 16.81 & 0.04 \\
 & 54601.5 & 16.91 & 0.04 \\
 & 54608.2 & 17.37 & 0.04 \\
 & 54616.2 & 18.01 & 0.12 \\
 & 54617.9 & 18.16 & 0.07 \\
 & 54624.2 & 18.54 & 0.23 \\
 & 54634.9 & 19.48 & 0.13 \\
 & 54643.6 & 19.74 & 0.16 \\
 & 54651.5 & 20.40 & 0.27 \\
 & 54665.0 & 20.76 & 0.30 \\
$uvw1$ & 54599.8 & 16.91 & 0.04 \\
 & 54601.5 & 16.90 & 0.03 \\
 & 54608.2 & 17.30 & 0.04 \\
 & 54616.1 & 17.99 & 0.10 \\
 & 54617.9 & 17.78 & 0.05 \\
 & 54624.1 & 18.16 & 0.11 \\
 & 54634.9 & 19.04 & 0.10 \\
 & 54643.6 & 19.97 & 0.21 \\
 & 54651.4 & 20.23 & 0.28 \\
 & 54664.9 & 20.90 & 0.40 \\
$u$ & 54599.8 & 16.82 & 0.03 \\
 & 54601.5 & 16.81 & 0.03 \\
 & 54608.2 & 17.05 & 0.03 \\
 & 54616.1 & 17.47 & 0.08 \\
 & 54617.9 & 17.41 & 0.04 \\
 & 54624.1 & 18.09 & 0.10 \\
 & 54634.9 & 18.48 & 0.08 \\
 & 54643.6 & 19.10 & 0.13 \\
 & 54651.4 & 19.19 & 0.16 \\
 & 54664.9 & 20.12 & 0.31 \\
$B$ & 54599.8 & 17.81 & 0.04 \\
 & 54601.5 & 17.79 & 0.04 \\
 & 54608.2 & 17.97 & 0.04 \\
 & 54616.1 & 18.23 & 0.09 \\
 & 54617.9 & 18.23 & 0.05 \\
 & 54624.1 & 18.45 & 0.09 \\
 & 54634.9 & 19.18 & 0.08 \\
 & 54643.6 & 19.28 & 0.11 \\
 & 54651.5 & 20.09 & 0.26 \\
 & 54664.9 & 20.08 & 0.20 \\
 & 54698.9 & 21.37 & 0.21 \\
$V$ & 54599.8 & 17.80 & 0.06 \\
 & 54601.5 & 17.70 & 0.06 \\
 & 54608.2 & 17.96 & 0.07 \\
 & 54616.2 & 17.96 & 0.14 \\
 & 54617.9 & 17.94 & 0.07 \\
 & 54624.2 & 17.94 & 0.16 \\
 & 54634.9 & 18.54 & 0.10 \\
 & 54643.6 & 18.66 & 0.12 \\
 & 54651.5 & 19.33 & 0.23 \\
 & 54665.0 & 19.31 & 0.23 \\
 & 54681.6 & 20.44 & 0.24 \\
 & 54696.5 & 20.43 & 0.16 \\
 & 54710.8 & 21.08 & 0.21 \\
\enddata
\tablenotetext{a}{Days in JD-2,400,000.5}
\tablenotetext{b}{Magnitudes corrected for a Galactic extinction of E($B-V$)=0.012 mag (Schlegel \etal 1998).}
\end{deluxetable}

\clearpage

\end{document}